\documentclass[aps,prl,twocolumn,amssymb,amsmath,nobibnotes,showpacs,
showkeys,superscriptaddress,floatfix]{revtex4}
\usepackage{graphicx}
\usepackage{dcolumn}
\usepackage{bm}
\usepackage{color}
\begin{document}
\title{Dynamical phase transitions in long-range Hamiltonian systems and Tsallis
distributions with a time-dependent index}
\author{Alessandro Campa}
\affiliation{Theoretical Physics Unit, Health and Technology Department,
Istituto Superiore di Sanit\`a, and INFN Roma1, Gruppo Collegato Sanit\`a,
Viale Regina Elena 299, 00161 Roma, Italy}
\author{Pierre-Henri Chavanis}
\affiliation{Laboratoire de Physique Th\'eorique - IRSAMC, CNRS
 Universit\'e Paul Sabatier, 31062 Toulouse, France }%
\author{Andrea Giansanti}
\affiliation{Physics Department, Universit\`a di Roma ``La Sapienza'', 
Piazzale Aldo Moro 2, 00185 Roma, Italy}
\author{Gianluca Morelli}
\affiliation{Physics Department, Universit\`a di Roma ``La Sapienza'' 
Piazzale Aldo Moro 2, 00185 Roma, Italy}
\date{\today}

\begin{abstract}
We study dynamical phase transitions in systems with long-range
interactions, using the Hamiltonian Mean Field (HMF) model as a simple
example. These systems generically undergo a violent relaxation to a
quasi-stationary state (QSS) before relaxing towards Boltzmann
equilibrium. In the collisional regime, the out-of-equilibrium
one-particle distribution function (DF) is a quasi-stationary solution
of the Vlasov equation, slowly evolving in time due to finite $N$
effects. For subcritical energy densities, we exhibit cases
where the DF is well-fitted by a Tsallis $q$-distribution with an
index $q(t)$ slowly decreasing in time from $q\simeq 3$ (semi-ellipse)
to $q=1$ (Boltzmann). When the index $q(t)$ reaches an energy dependent
critical value $q_{crit}$, the non-magnetized (homogeneous) phase becomes Vlasov
unstable and a dynamical phase transition is triggered, leading to a
magnetized (inhomogeneous) state. While Tsallis
distributions play an important role in our study, we explain this
dynamical phase transition by using only conventional statistical
mechanics. For supercritical energy densities, we report for the first time
the existence of a magnetized QSS with a very long lifetime.
\end{abstract}

\maketitle

Systems with long-range interactions have a peculiar dynamics and
thermodynamics \cite{houches}. In these systems, the potential
of interaction decreases at large distances like $r^{-\alpha}$, with an
exponent $\alpha$ smaller than the dimension of space. As a result, the energy
is non-additive and this can lead to
negative specific heats and inequivalence of statistical ensembles.
The equilibrium properties of these systems are now well-understood
and the main challenge concerns the description of their
out-of-equilibrium properties. The general methodology to study these
systems can have impact in different domains of physics (see various
contributions in \cite{houches}).

Long-range systems can spontaneously develop out-of-equilibrium coherent structures,
called quasi stationary states (QSS). The QSS are generically
characterized by non-Boltzmannian distributions and their lifetime
diverges with the size $N$ of the system.
Galaxies in the universe and 2D vortices in geophysical and
astrophysical flows are examples of such QSS \cite{csr}.
They are also observed in non-neutral plasmas under a
strong magnetic field \cite{hd}, in the free electron laser
\cite{barre} and in simple models like the Hamiltonian mean-field
(HMF) model
\cite{latora,yama,cvb,epjb,anto,campa}. They are
therefore general features of systems with long-range
interactions. The interesting issues about the nature
of these QSSs concern: i) their rapid formation starting
from a generic non equilibrium initial condition; ii) their structure
and stability; iii) their slow evolution due to finite $N$
effects; iv) their final transition to the Boltzmann equilibrium
distribution. In particular, point iv) requires to develop a theory of
{\it dynamical phase transition} adapted to these systems.

In this perspective, the HMF model \cite{ruffo} became a paradigmatic
laboratory. It represents a system of $N$ globally coupled rotators
of unit mass described by the Hamiltonian:
\begin{equation}\label{ham}
H = \frac{1}{2}\sum_{i=1}^N p_i^2 + 
\frac{1}{2N} \sum_{i,j=1}^{N} \left[1 - \cos(\theta_i - \theta_j )\right]\, ,
\end{equation}
where $\theta_i$ represents the orientation of the $i$-th rotator and
$p_i$ is the conjugated momentum. The evolution of the system
may be followed through the magnetization, an order parameter defined
as $M=|{\mathbf M}|=|\sum {\mathbf m_i}| /N$, where
${\mathbf m_i}=(\cos \theta_i,\sin \theta_i)$. The
HMF model has been extensively studied as a representative of a broad
class of systems with long-range interactions, like gravitational
systems (see references in \cite{cvb}). The equilibrium solution \cite{ruffo}
reveals the existence of a
second-order phase transition at the critical energy density
$U_{c}=3/4$: the Boltzmann equilibrium state is homogeneous
(non-magnetized) for $U>U_c$ and inhomogeneous (magnetized) for
$U<U_c$.

On the other hand, the interpretation of QSS is an issue of tense
debate that has recently polarized researchers in two groups. The
first group \cite{latora} interprets the QSS as non-Boltzmannian equilibrium
states explainable within Tsallis generalized thermodynamics
\cite{tsallis} when the thermodynamic limit $N\rightarrow +\infty$ is
taken before the infinite time limit $t\rightarrow +\infty$. The
justification advocated is that the evolution of the system is
generally non-ergodic so that non-standard distributions can emerge.
The other group \cite{yama,cvb,epjb,anto} interprets the QSS in terms
of the Lynden-Bell \cite{lb} theory of violent relaxation that was
initially developed for collisionless stellar systems and 2D vortices
\cite{csr}. For sufficiently short times, or for $N\rightarrow
+\infty$, the evolution of the one-particle distribution function (DF)
is governed by the Vlasov equation, a mean field equation that ignores
correlations between particles \cite{hb3}. In this context, the QSS
result from a process of phase mixing and violent relaxation
on a coarse-grained scale. The statistical theory of
Lynden-Bell \cite{lb} predicts the {\it most mixed state} consistent
with all the constraints of the collisionless (Vlasov) dynamics.

Numerical simulations have given contrasting results. For a water-bag
initial condition (i.e., uniform distribution of velocities and angles
in an interval symmetric around $0$) with energy density $U=0.69$, it has been
found that the Lynden-Bell theory predictions are correct
\cite{anto} if the initial value of the magnetization $M_0\equiv
M(t=0)$ is below a critical value $(M_{0})_{crit}=0.897$ appearing in
the theory \cite{epjb,anto}, while this is no more true if $M_0$ is close
to $1$ \cite{latora,campa}. In particular, for $M_0=1$, Campa
{\it et al.} \cite{campa} find that the QSS is well-fitted by a homogeneous
($M_{QSS}=0$) semi-elliptical distribution, which is a particular Tsallis DF
with $q=3$ (as noted in \cite{hb3}), while the Lynden-Bell
prediction is an inhomogeneous ($M_{QSS}\neq 0$) Boltzmann distribution
corresponding to the non degenerate limit of the
theory \cite{hb3,note}. 
Stimulated by these results, one of us
\cite{epjb} has proposed an approach in which the two points of view
can be reconciled in the following manner: (i) if the system mixes
well (ergodicity), the QSS is (close to) the ``most
mixed state'' predicted by Lynden-Bell's theory; (ii) if the system
does not mix well (non ergodicity), relaxation is {\it
incomplete} and non-standard DFs (including the
important class of Tsallis distributions) can occur.

The  purpose of this Rapid Communication is to give numerical evidence that
Tsallis distributions can also characterize the slow
collisional evolution of the QSS (due to finite $N$ effects) up to the relaxation
to Boltzmann equilibrium. Our results support the proposal made in
Ref. \cite{hb3}, according to which the out-of-equilibrium DF, for a
large class of initial conditions, could be fitted by a Tsallis
$q$-distribution (see below) with an index $q(t)$ depending on
time. We will show that the dynamical phase transition
from the non-magnetized QSS to the magnetized Boltzmann distribution is triggered
when the value $q(t)$ is such that the DF becomes Vlasov
unstable: therefore, the final transition to the
Boltzmann distribution is driven by a dynamical instability with
respect to the Vlasov equation.

In a non-magnetized QSS, the DF depends only on
velocity and slowly evolves in time due to finite $N$ effects. It
turns out that the time-dependent DF is very well fitted by Tsallis distributions
(polytropes):
\begin{eqnarray}
\label{e1}
f(v,t)=C(t)\left\lbrack 1-\left (\frac{v}{v_{max}(t)}\right )^2\right \rbrack^{1/(q(t)-1)},
\end{eqnarray}
with a time dependent index $q(t)\ge 1$. The support of the function
is the interval $[-v_{max}(t),v_{max}(t)]$; the normalization factor
$C(t)$ and $v_{max}(t)$ depend on time through $q(t)$. The normalization
condition $\int_{-\infty}^{+\infty}fdv=1/(2\pi)$ and the conservation of energy
$U-1/2=\pi\int_{-\infty}^{+\infty}fv^2 dv$ lead to the relations
$1/(2\pi)=C v_{max} \sqrt{\pi}\Gamma(1+1/(q-1))/\Gamma(3/2+1/(q-1))$
and $U-1/2=\pi C v_{max}^3
(\sqrt{\pi}/2)\Gamma(1+1/(q-1))/\{(\Gamma(3/2+1/(q-1))(3/2+1/(q-1))\}$ from
which we obtain the identity
$v_{max}^{2}(t)=4(U-1/2)(3/2+1/(q(t)-1))$.
On the other hand, using the stability criterion
$1+\pi\int_{-\infty}^{+\infty}(f'(v)/v) dv\ge 0$ of \cite{inagaki,yama}
(see also \cite{cvb}) we find that a spatially homogeneous $q$-distribution is
Vlasov stable (dynamical stability) iff $U\ge U_{crit}(q)\equiv
1/2+(1/4)(q+1)/(3q-1)$ \cite{av,cvb,hb3}.
For $q=1$ (Boltzmann) we recover $U_{crit}(1)=U_{c}=3/4$, for
$q\rightarrow +\infty$ (water-bag) we recover
$U_* \equiv U_{crit}(\infty)=7/12$ and for
$q=3$ (semi-elliptical),
we recover $U_{crit}(3)=5/8$ \cite{campa}. From this
criterion, DFs of the form (\ref{e1}) are always stable for $U>U_c=3/4$
and always unstable for $U<U_*=7/12$. For $U_*<U<U_c$, they are
stable iff $q>q_{crit}(U)$ with \cite{note2}
\begin{eqnarray}
\label{e2}
q_{crit}(U)=\frac{4U-1}{12U-7}.
\end{eqnarray}

Campa {\it et al.} \cite{campa} considered a
water-bag initial condition with $M_0=0$ at energy density
$U=0.69$. This initial DF is already a maximum Lynden-Bell entropy
state so it does not experience phase mixing and violent relaxation
\cite{hb3}.  However, it slowly evolves due to finite $N$ effects.
In Ref. \cite{campa} it was found
that the system rapidly forms a velocity DF of
semi-elliptical shape. As we have already indicated, this is a Tsallis
DF of the form (\ref{e1}) with $q=3$. However, the slow evolution of the
DF on longer times was not quantitatively addressed. Here, we have performed
simulations at several energies, all of
them starting from an unmagnetized state (uniform angle distribution).
For the subcritical energy densities $U<U_c$, the initial velocity
distribution was a semi-elliptical, $q=3$, function. The energy densities were
all above the critical value $5/8$ corresponding to $q=3$, and
therefore the initial steady states were Vlasov stable. In these cases
we found that during the QSS, while the magnetization
remains zero (or, more precisely, oscillates about a value scaling to zero as
$1/\sqrt{N}$ because of finite size effects), the velocity
DF has a slow evolution. At a certain point the
distribution is no more Vlasov stable; then the magnetization begins
to increase relatively fast, and the systems heads towards Boltzmann
equilibrium. During the slow process characterizing the QSS, we found
that the velocity DF remains always close to one of
the $q$ functions (\ref{e1}), so that it is possible to define a
function of time $q(t)$. In this way, the evolution of the velocity
DF is described by the slow decay of a {\it single}
quantity, the index $q(t)$ of the polytrope, from the value $q(0)=3$
to the energy dependent critical value $q_{crit}(U)$.

In Fig. 1, we show results corresponding to $U=0.69$ (similar results,
not shown, were obtained for $U=0.65$, $U=0.67$, and $U=0.71$). The
left panel shows the evolution of the magnetization $M(t)$, while the
right panel shows the evolution of the function $q(t)$.  The runs
correspond to $N=2^{14}$. The critical value of $q$ corresponding to
$U=0.69$ is $q_{crit}(0.69)=1.375$. This value is indicated, for
reference, as a dashed line in the graph of $q(t)$.
At each time, the best value of $q$ is determined by a least square
procedure, minimizing with respect to $q$ the sum of the squares of
the differences between the numerical distribution and the
$q$-distribution for the energy density $U=0.69$. In this procedure, both the
numerical and the theoretical distributions are given as histograms
with a suitably chosen bin width; $q$ is varied in steps of
$0.001$. We see that the destabilization of the DF, and
the associated process of magnetization of the system, is related to
the approach of $q(t)$ towards the critical value $q_{crit}(U)$. We
found that this behavior is present for all $N$ values that we have
studied, although noise obviously increases if $N$ is decreased. 

\begin{figure}[htbp!]
\begin{center}
\includegraphics[bb=60 300 510 510,clip,scale=0.54]{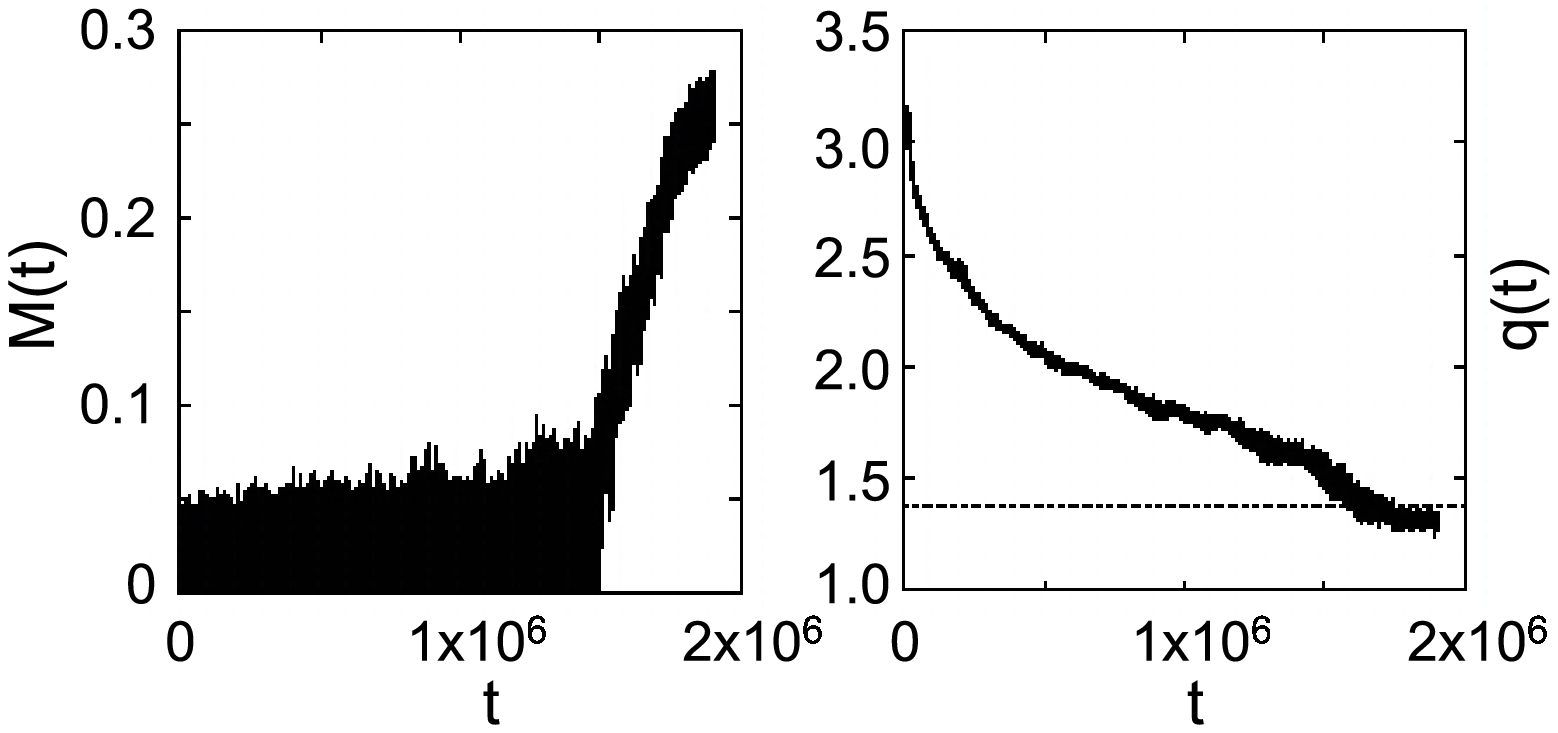}
\caption[]
{Time evolution of the magnetization $M(t)$ (left panel) and of the
index $q(t)$ (right panel), for $U=0.69$ and $N=2^{14}$. The
dotted line in the right panel indicates the critical value
$q_{crit}(0.69)=1.375$. During the collisional
process, the DF is a quasi-stationary solution of the Vlasov equation
of the form (\ref{e1}) slowly evolving in time due to finite $N$
effects (see Figure 2). When $q(t)$ reaches $q_{crit}$, it becomes
dynamically (Vlasov) unstable and the phase transition is triggered.}
\end{center}
\end{figure}

In Fig. 2 we show snapshots of the DF during the ``collisional''
relaxation for the same run as that shown in the previous
figure. The full lines give the numerical distributions at a given
time, and the dotted lines are the $q$-functions for the best fitted
value of $q(t)$. 
The fit by a Tsallis DF with a time dependent index is
very good during all the out-of-equilibrium evolution.
\begin{figure}[htbp!]
\begin{center}
\includegraphics[bb=80 175 510 595,clip,scale=0.54]{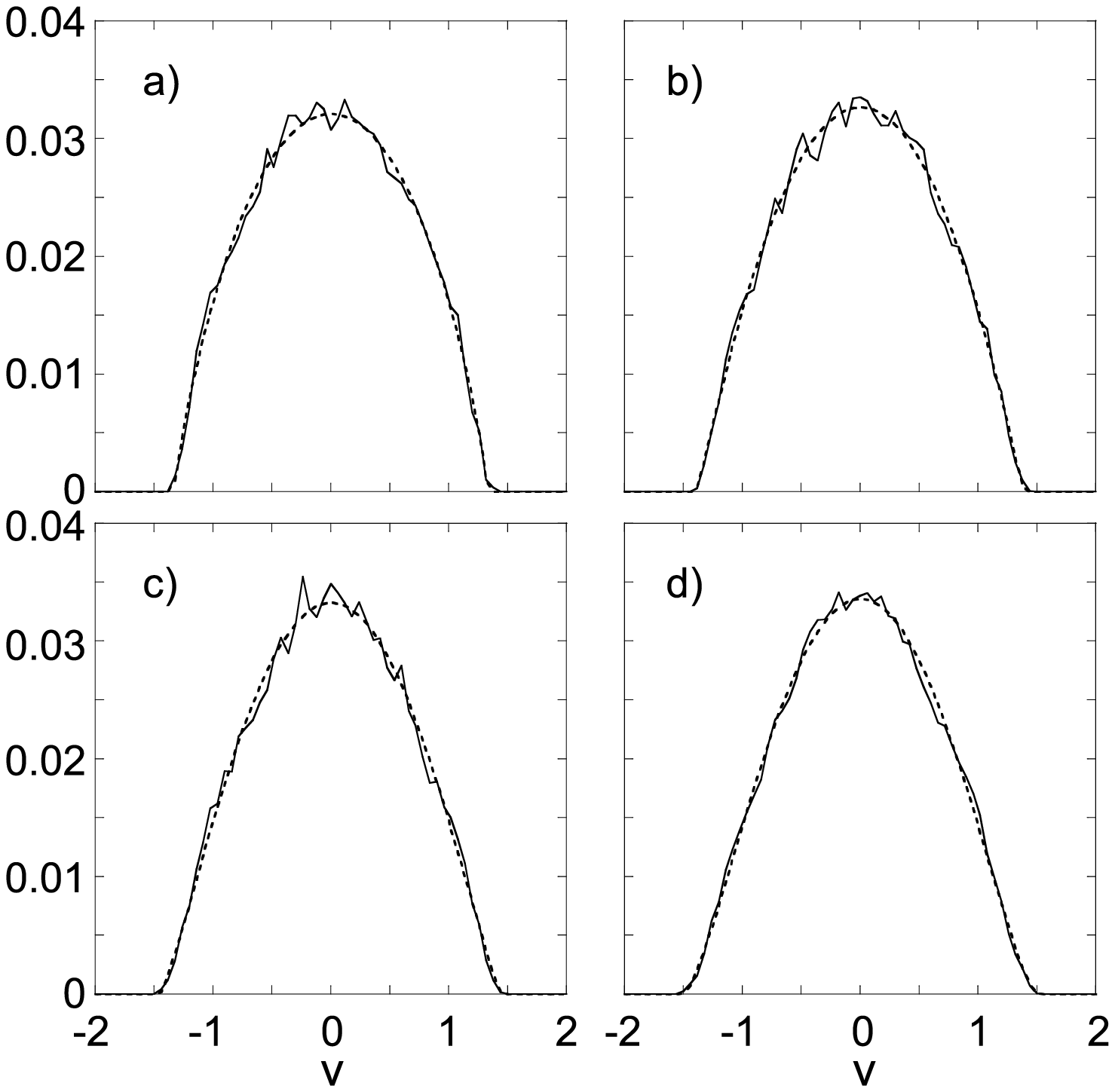}
\caption[]
{Velocity DF at four different times during the QSS of the run shown in
Figure 1. The full lines are the numerical distributions, while the dotted
lines are the $q$-functions for the best fitted value of $q(t)$.
a) $t=3\times 10^5$ and $q(t)=2.224$; b) $t=6\times 10^5$ and $q(t)=1.987$;
c) $t=9\times 10^5$ and $q(t)=1.789$; d) $t=1.2\times 10^6$ and $q(t)=1.697$.
The absolute value of the DF depends on the chosen bin width. }
\end{center}
\end{figure}

Finally, we consider the case of a supercritical energy density $U>U_c=3/4$.
For $U=0.8$, Campa {\it et al.} \cite{campa} started from a
non-magnetized state with a semi-elliptical distribution ($q=3$),
which is Vlasov stable, and found that the system relaxes to the
Boltzmann equilibrium distribution on a very slow timescale scaling
like $t_{coll}(N)\sim e^{N}$. Since $q_{crit}(U)<1$ for $U>U_c$, while
Tsallis distributions with a compact support have $q(t)>1$, the
spatially homogeneous phase remains always Vlasov stable during the
collisional evolution and no dynamical phase transition occurs. This
explains \cite{hb3} the long observed relaxation time for $U>U_{c}$ as
compared to the case $U<U_{c}$. In this case, the relaxation is due to
a slow collisional relaxation, not to a dynamical (Vlasov)
instability.
Considering a slightly different supercritical energy density,
$U=0.85$, for which $q_{crit}=0.75$, the left panel of Fig. 3 shows the
function $q(t)$ that, starting from $q(0)=3$, slowly approaches $1$.
During this relaxation the DF is well
approximated, as in the subcritical case shown in Fig. 2,
by the $q$-function with $q=q(t)$.
We have also studied another type of initial condition,
at the supercritical energy density $U=0.85$, starting from
a non-magnetized state with a Tsallis velocity DF
corresponding to $q=0.5$, which is below the critical value
$0.75$.  This initial DF is Vlasov unstable, linear
instability followed by phase mixing and violent
relaxation quickly drive the system towards a magnetized
out-of-equilibrium QSS. Although the system should eventually relax to
a non-magnetized Boltzmann distribution, the right panel of Fig. 3, corresponding
to a run with $N=2^{11}$, shows that the duration of this magnetized QSS is
extraordinary long (probably exponential in $N$, as estimated by runs
at different $N$ values).  For comparison, we note that for $N=2^{11}$
the QSS at $U=0.69$ lasts for a time of about $5\times 10^4$
\cite{campa}.

In conclusion, we have shown that Tsallis $q$-distributions
(polytropes) can be useful to describe certain features of the complex
HMF dynamics. First, they can provide good fits of the QSS in certain
cases of incomplete relaxation \cite{campa,hb3}. However, this is not
general \cite{hb3} and other DF that are stable stationary solutions
of the Vlasov equation can be observed as well (the system can also
have a non-stationary oscillatory behaviour \cite{mk}). More
interestingly, like in the gravitational simulations of Taruya and
Sakagami \cite{ts}, we have found that the out-of-equilibrium
evolution of the DF in the collisional regime can be fitted by Tsallis
distributions with a time dependent index $q(t)$. This may be a general
feature of the dynamics of long-range systems.
However, while Taruya and Sakagami \cite{ts}
interpret the dynamical phase transition as a generalized
thermodynamical instability (in Tsallis sense), we interpret it as a
dynamical instability with respect to the Vlasov equation
\cite{cst,hb3}.
\begin{figure}[htbp!]
\begin{center}
\includegraphics[bb=60 300 510 510,clip,scale=0.54]{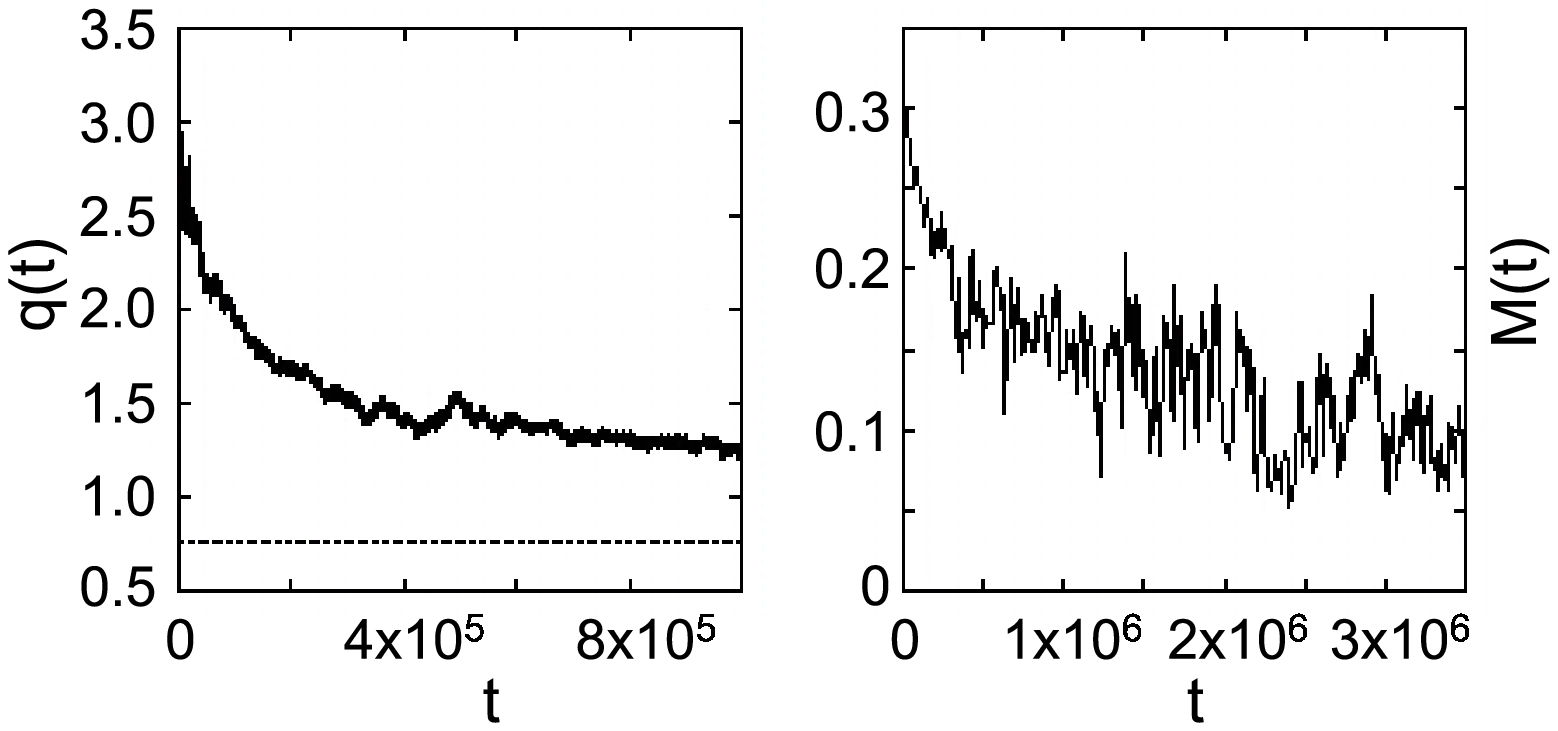}
\caption[]
{Time evolution of the index $q(t)$ (left panel) and of the magnetization
$M(t)$ (right panel), for $U=0.85$ and $N=2^{11}$. The left panel refers to a
run with $q(0)=3$, in which the system remains always demagnetized ($q(t)>q_{crit}=0.75$) and slowly
approaches Boltzmann equilibrium ($q=1$); the right panel refers to a run with $q(0)=0.5<q_{crit}=0.75$,
in which the QSS is magnetized and lasts for a very long time.}
\end{center}
\end{figure}

The evolution of the system from the QSS to the
Boltzmann distribution, due to the development of correlations
(``collisions'') between particles, could be understood by developing a
kinetic theory of the Hamiltonian $N$-body problem. As a matter of fact,
starting from the
Liouville equation and using an expansion of the solutions of the
BBGKY-like hierarchy in powers of $1/N$ in a proper thermodynamic
limit $N\rightarrow +\infty$, a general kinetic equation can be
obtained at the order $O(1/N)$ \cite{hb3}; the BBGKY
hierarchy is closed, neglecting three body and higher correlations
of order $1/N^2$ or smaller. However, for one dimensional systems
like the HMF model, this kinetic equation does not tend to the
Boltzmann distribution (for spatially homogeneous
systems the collision term, which reduces to the Landau or
Lenard-Balescu operator \cite{bd,hb3}, cancels out identically in
$1D$). This implies that higher order correlations are needed to
understand the convergence of the system to the Boltzmann
distribution. Therefore, we can only conclude that the collisional
relaxation time $t_{coll}(N)$ is larger than $Nt_{D}$ (where $t_{D}$
is a typical dynamical time of the system), but its precise scaling
with $N$ cannot be deduced from the present kinetic
theory. Different scalings have been found numerically
depending on the initial condition: $Nt_{D}$
\cite{latora}, $N^{1.7}t_{D}$ \cite{yama,campa} and $e^{N}t_{D}$
\cite{campa}. Due to the considerable timescale separation between
the dynamical time (of order $t_D$) and the collision time (larger
than $Nt_D$), it can be argued that, during the slow ``collisional''
relaxation, the time dependent DF remains close to a steady solution
of the Vlasov equation that slowly evolves under the effect of
collisions (finite $N$ effects). In this Rapid Communication we have shown
that, at least for a class of initial conditions, these slowly
evolving QSS  are well-fitted by Tsallis
$q(t)$-functions. However, as discussed in Sec. 5 of
\cite{cst} this is not necessarily the mark of a ``generalized
thermodynamics'' \cite{tsallis}.
Indeed, if using only classical methods of kinetic
theory (consistent with the Boltzmann entropy), we were able to write
down the kinetic equation describing the collisional relaxation of the
spatially homogeneous HMF model (the analogue of the
orbit-averaged-Fokker-Planck equation in astrophysics), the DF
(\ref{e1}) could form an {\it approximate} time dependent solution of
this classical kinetic equation (without relation to generalized
thermodynamics). Unfortunately, for the HMF model, this kinetic
equation is not known.


\begin{thebibliography}{}


\bibitem{houches}
{\it Dynamics and thermodynamics of systems
with long range interactions}, edited by T. Dauxois {\it et al.}, Lecture
Notes in Physics {\bf 602} (Springer, 2002); {\it Dynamics and thermodynamics of systems
with long range interactions: Theory and experiments}, edited by A. Campa
{\it et al.}, AIP Conf. Proc. {\bf 970} (AIP, 2008).
\bibitem{csr}
P.H. Chavanis {\it et al.}, Astrophys. J.  {\bf 471}, 385 (1996).
\bibitem{hd}
X. P. Huang and C. F. Driscoll, Phys. Rev. Lett. {\bf 72}, 2187 (1994).
\bibitem{barre}
J. Barr\'e {\it et al.}, Phys. Rev. E {\bf 69}, 045501 (2004).
\bibitem{latora}
V. Latora {\it et al.}, Phys. Rev. E  {\bf 64}, 056134 (2001).
\bibitem{yama}
Y. Yamaguchi {\it et al.}, Physica A {\bf 337}, 36 (2004).
\bibitem{cvb}
P.H. Chavanis {\it et al.}, Eur. Phys. J. B {\bf 46}, 61 (2005).
\bibitem{epjb}
P.H. Chavanis, Eur. Phys. J. B {\bf 53}, 487 (2006).
\bibitem{anto}
A. Antoniazzi {\it et al.}, Phys. Rev. E {\bf 75}, 011112 (2007).
\bibitem{campa}
A. Campa {\it et al.}, Phys. Rev. E {\bf 76}, 041117 (2007).
\bibitem{ruffo}
M. Antoni and S. Ruffo, Phys. Rev. E {\bf 52}, 2361 (1995).
\bibitem{tsallis}
C. Tsallis, J. Stat. Phys. {\bf  52}, 479 (1988).
\bibitem{lb}
D. Lynden-Bell, Mon. Not. R. Astron. Soc. {\bf 136}, 101 (1967).
\bibitem{hb3}
P.H. Chavanis, Physica A {\bf 387}, 787 (2008).
\bibitem{note}
As pointed out  in \cite{epjb,hb3}, the failure of the Lynden-Bell prediction for
$M_0=1$ may be due to the proximity of this value to the critical
magnetization $(M_{0})_{crit}=0.897$, very close to $1$ for the energy
density $U=0.69$.
\bibitem{inagaki}
S. Inagaki and T. Konishi, Publ. Astron. Soc. Jpn.  {\bf 45}, 733 (1993).
\bibitem{av}
C. Anteneodo and R. Vallejos, Physica A  {\bf 344}, 383 (2004).
\bibitem{note2}
Below this critical value, the DF (2) ceases to be a
maximum of the Casimir functional $S_q=-1/(q-1)\int (f^q-f)
{\rm d}\theta {\rm d}v$ (at fixed mass and energy) and it becomes a
saddle point \cite{yama,cvb}.
\bibitem{mk}
H. Morita and K. Kaneko, Phys. Rev. Lett. {\bf 96}, 050602 (2006).
\bibitem{ts}
A. Taruya and M. Sakagami, Phys. Rev. Lett. {\bf 90}, 181101 (2003).
\bibitem{cst}
P.H. Chavanis and C. Sire, Physica A  {\bf 356}, 419 (2005).
\bibitem{bd}
F. Bouchet, T. Dauxois, Phys. Rev. E  {\bf 72}, 045103 (2005).


\end{thebibliography}
\end{document}